\documentclass[12pt,titlepage]{article}
\usepackage{amssymb}

\input{tcilatex}

\begin{document}

\title{BOUND STATES BY A PSEUDOSCALAR COULOMB POTENTIAL IN ONE-PLUS-ONE DIMENSIONS}
\date{}
\author{Antonio S. de Castro \\
%EndAName
\\
UNESP - Campus de Guaratinguet\'{a}\\
Departamento de F\'{\i}sica e Qu\'{\i}mica\\
Caixa Postal 205\\
12516-410 Guaratinguet\'{a} SP - Brasil\\
\\
Electronic mail: castro@feg.unesp.br}
\maketitle

\begin{abstract}
The Dirac equation is solved for a pseudoscalar Coulomb potential in a
two-dimensional world. An infinite sequence of bounded solutions are
obtained. These results are in sharp contrast with those ones obtained in
3+1 dimensions where no bound-state solutions are found. Next the general
two-dimensional problem for pseudoscalar power-law potentials is addressed
consenting us to conclude that a nonsingular potential leads to bounded
solutions. The behaviour of the upper and lower components of the Dirac
spinor for a confining linear potential nonconserving- as well as
conserving-parity, even if the potential is unbounded from below, is
discussed in some detail.
\end{abstract}

The Coulomb potential of a point electric charge in a 1+1 dimension,
considered as the time component of a Lorentz vector, is linear and so it
provides a constant electric field always pointing to, or from, the point
charge. This problem is related to the confinement of fermions in the
Schwinger and in the massive Schwinger models \cite{col1}-\cite{col2} and in
the Thirring-Schwinger model \cite{fro}. It is frustrating that, due to the
tunneling effect (Klein\'{}s paradox), there are no bound states for this
kind of potential regardless of the strength of the potential \cite{cap}-%
\cite{gal}. The linear potential, considered as a Lorentz scalar, is also
related to the quarkonium model in one-plus-one dimensions \cite{hoo}-\cite
{kog}. Recently it was incorrectly concluded that even in this case there is
solely one bound state \cite{bha}. Later, the proper solutions for this last
problem were found \cite{cas}-\cite{hil}. However, it is well known from the
quarkonium phenomenology in the real 3+1 dimensional world that the best fit
for meson spectroscopy is found for a convenient mixture of vector and
scalar potentials put by hand in the equations (see, \textit{e.g.}, \cite
{luc}). The mixed vector-scalar linear potential in 1+1 dimensions was
recently considered \cite{asc1}. There it was found that there are
analytical bound-state solutions on condition that the scalar component of
the potential is of sufficient strength compared to the vector component ($%
|V_{s}|\geq |V_{t}|$). As a by-product, that approach also showed that there
exist relativistic confining potentials providing no bound-state solutions
in the nonrelativistic limit. Although the discussion was confined to the
vector-scalar mixing, the inclusion of a pseudoscalar potential could also
be allowed.

In a recent paper, McKeon and Van Leeuwen \cite{mck} considered a
non\-con\-ser\-ving-parity pseu\-do\-sca\-lar Coulomb (NCPPC) potential ($%
V=\lambda /r$) in 3+1 dimensions and concluded that there are no bounded
solutions for the reason that the different parity eigenstates mix.
Furthermore, they asserted that \textit{the absence of bound states in this
system confuses the role of the }$\pi $\textit{-meson in the binding of
nucleons}. Such an intriguing conclusion sets the stage for the analyses by
other sorts of pseudoscalar potentials. A natural question to ask is if the
absence of bounded solutions by a NCPPC potential is a characteristic
feature of the four-dimensional world. With this in mind, we approach in the
present paper the less realistic Dirac equation in one-plus-one dimensions
with a NCPPC potential ($V=m\omega |x|$).

The two-dimensional Dirac equation can be obtained from the
four-dimen\-sional one with the mixture of spherically symmetric scalar,
vector and anomalous magnetic-like (tensor) interactions. If we limit the
fermion to move in the $x$-direction ($p_{y}=p_{z}=0$) the four-dimensional
Dirac equation decomposes into two equivalent two-dimensional equations with
2-component spinors and 2$\times $2 matrices \cite{str}. Then, there results
that the scalar and vector interactions preserve their Lorentz structures
whereas the anomalous magnetic interaction turns out to be a pseudoscalar
interaction. Furthermore, in the 1+1 world there is no angular momentum so
that the spin is absent. Therefore, the 1+1 dimensional Dirac equation allow
us to explore the physical consequences of the negative-energy states in a
mathematically simpler and more physically transparent way. The confinement
of fermions by a pure \-con\-ser\-ving-parity pseudoscalar double-step
potential \cite{asc2} and their scattering by a pure
non\-con\-ser\-ving-parity pseudoscalar step potential \cite{asc3} have
already been analyzed in the literature providing the opportunity to find
some quite interesting results. Indeed, the two-dimensional version of the
anomalous magnetic-like interaction linear in the radial coordinate,
christened by Moshinsky and Szczepaniak \cite{ms} as Dirac oscillator, has
also received attention. Nogami and Toyama \cite{nt}, Toyama \textit{et al.} 
\cite{tplus} and Toyama and Nogami \cite{tn} studied the behaviour of wave
packets under the influence of that conserving-parity potential whereas
Szmytkowski and Gruchowski \cite{sg} proved the completeness of the
eigenfunctions. More recently Pacheco \textit{et al.} \cite{pa} studied some
thermodynamics properties of the 1+1 dimensional Dirac oscillator.

Let us begin by presenting the Dirac equation in 1+1 dimensions. In the
presence of a time-independent potential the 1+1 dimensional
time-independent Dirac equation for a fermion of rest mass $m$ reads 
\begin{equation}
\mathcal{H}\Psi =E\Psi  \label{eq1}
\end{equation}

\begin{equation}
\mathcal{H}=c\alpha p+\beta mc^{2}+\mathcal{V}  \label{eq1a}
\end{equation}

\noindent where $E$ is the energy of the fermion, $c$ is the velocity of
light and $p$ is the momentum operator. $\alpha $ and $\beta $ are Hermitian
square matrices satisfying the relations $\alpha ^{2}=\beta ^{2}=1$, $%
\left\{ \alpha ,\beta \right\} =0$. From the last two relations it steams
that both $\alpha $ and $\beta $ are traceless and have eigenvalues equal to 
$\pm $1, so that one can conclude that $\alpha $ and $\beta $ are
even-dimensional matrices. One can choose the 2$\times $2 Pauli matrices
satisfying the same algebra as $\alpha $ and $\beta $, resulting in a
2-component spinor $\Psi $. The positive definite function $|\Psi |^{2}=\Psi
^{\dagger }\Psi $, satisfying a continuity equation, is interpreted as a
probability position density and its norm is a constant of motion. This
interpretation is completely satisfactory for single-particle states \cite
{tha}. We use $\alpha =\sigma _{1}$ and $\beta =\sigma _{3}$. For the
potential matrix we consider 
\begin{equation}
\mathcal{V}=1V_{t}+\beta V_{s}+\alpha V_{e}+\beta \gamma ^{5}V_{p}
\label{eq2}
\end{equation}

\noindent where $1$ stands for the 2$\times $2 identity matrix and $\beta
\gamma ^{5}=\sigma _{2}$. This is the most general combination of Lorentz
structures for the potential matrix because there are only four linearly
independent 2$\times $2 matrices. The subscripts for the terms of potential
denote their properties under a Lorentz transformation: $t$ and $e$ for the
time and space components of the 2-vector potential, $s$ and $p$ for the
scalar and pseudoscalar terms, respectively. It is worth to note that the
Dirac equation is covariant under $x\rightarrow -x$ if $V_{e}(x)$ and $%
V_{p}(x)$ change sign whereas $V_{t}(x)$ and $V_{s}(x)$ remain the same.
This is because the parity operator $P=\exp (i\varepsilon )P_{0}\sigma _{3}$%
, where $\varepsilon $ is a constant phase and $P_{0}$ changes $x$ into $-x$%
, changes sign of $\alpha $ and $\beta \gamma ^{5}$ but not of $1$ and $%
\beta $.

Defining the spinor $\psi $ as 
\begin{equation}
\psi =\exp \left( \frac{i}{\hslash }\Lambda \right) \Psi  \label{eq5}
\end{equation}

\noindent where 
\begin{equation}
\Lambda (x)=\int^{x}dx^{\prime }\frac{V_{e}(x^{\prime })}{c}  \label{eq6}
\end{equation}

\noindent the space component of the vector potential is gauged away

\begin{equation}
\left( p+\frac{V_{e}}{c}\right) \Psi =\exp \left( \frac{i}{\hslash }\Lambda
\right) p\psi  \label{eq7}
\end{equation}

\noindent so that the time-independent Dirac equation can be rewritten as
follows:

\begin{equation}
H\psi =E\psi  \label{eq7a}
\end{equation}

\begin{equation}
H=\sigma _{1}cp+\sigma _{2}V_{p}+\sigma _{3}\left( mc^{2}+V_{s}\right)
+1V_{t}  \label{eq8}
\end{equation}

\noindent showing that the space component of a vector potential only
contributes to change the spinors by a local phase factor.

Provided that the spinor is written in terms of the upper and the lower
components 
\begin{equation}
\psi =\left( 
\begin{array}{c}
\phi \\ 
\chi
\end{array}
\right)  \label{eq8a}
\end{equation}

\noindent the Dirac equation decomposes into :

\begin{eqnarray}
\left( V_{t}-E+V_{s}+mc^{2}\right) \phi (x) &=&i\hslash c\chi ^{\prime
}(x)+iV_{p}\chi (x)  \nonumber \\
&&  \label{eq8b} \\
\left( V_{t}-E-V_{s}-mc^{2}\right) \chi (x) &=&i\hslash c\phi ^{\prime
}(x)-iV_{p}\phi (x)  \nonumber
\end{eqnarray}

\noindent where the prime denotes differentiation with respect to $x$. In
terms of $\phi $ and $\chi $ the spinor is normalized as $\int_{-\infty
}^{+\infty }dx\left( |\phi |^{2}+|\chi |^{2}\right) =1$, so that $\phi $ and 
$\chi $ are square integrable functions. It is clear from the pair of
coupled first-order differential equations (\ref{eq8b}) that both $\phi $
and $\chi $ must be discontinuous wherever the potential undergoes an
infinite jump and have opposite parities if the Dirac equation is covariant
under $x\rightarrow -x$. In the nonrelativistic approximation (potential
energies small compared to the rest mass) Eq. (\ref{eq8b}) loses all the
matrix structure and becomes

\begin{equation}
\chi =\frac{p}{2mc}\phi  \label{eq8c}
\end{equation}

\begin{equation}
\left( -\frac{\hslash ^{2}}{2m}\frac{d^{2}}{dx^{2}}+V_{t}+V_{s}+\frac{%
V_{p}^{2}}{2mc^{2}}+\frac{\hslash V_{p}^{\prime }}{2mc}\right) \phi =\left(
E-mc^{2}\right) \phi  \label{eq8d}
\end{equation}

\noindent Eq. (\ref{eq8c}) shows that $\chi $ if of order $v/c<<1$ relative
to $\phi $ and Eq. (\ref{eq8d}) shows that $\phi $ obeys the Schr\"{o}dinger
equation without distinguishing the contributions of vector and scalar
potentials (at this point the author digress to make his apologies for
mentioning in former papers (\cite{asc2}-\cite{asc3}) that the pseudoscalar
potential does not present any contributions in the nonrelativistic limit).
It is remarkable that the Dirac equation with a nonvector potential, or a
vector potential contaminated with some scalar or pseudoscalar coupling, is
not invariant under $V\rightarrow V+const.$, this is so because only the
vector potential couples to the positive-energies in the same way it couples
to the negative-ones, whereas nonvector contaminants couple to the mass of
the fermion. Therefore, if there is any nonvector coupling the absolute
values of the energy will have physical significance and the freedom to
choose a zero-energy will be lost. This last statement remains truthfully in
the nonrelativistic limit if one considers that such a contaminant is a
pseudoscalar potential.

Now, let us choose $V_{t}=V_{s}=0$ and the intrinsically relativistic NCPPC
potential $V_{p}=m\omega c|x|$, where $\omega $ is a real parameter.
Although $\omega <0$ gives rise to a potential unbounded from below, the
possibility of such a sort of potential to bind fermions is already
noticeable in the nonrelativistic limit of the theory (see Eq. (\ref{eq8d}%
)), where the pseudoscalar linear potential, whether it is a conserving- or
a nonconserving-parity potential, manifests itself effectively as a
quadratic potential. Defining

\begin{eqnarray}
\eta &=&\sqrt{\frac{2m|\omega |}{\hslash }}\,x  \label{eq8f1} \\
&&  \nonumber \\
\nu &=&\frac{E^{2}-m^{2}c^{4}}{2\hslash |\omega |mc^{2}}-\frac{1+s(\omega )}{%
2}  \label{eq8f}
\end{eqnarray}

\noindent where $s(\omega )$ stands for the sign function, \noindent
\noindent the Dirac equation (\ref{eq8b}) turns into the
Schr\"{o}dinger-like differential equations

\begin{eqnarray}
-\frac{d^{2}\phi }{d\eta ^{2}}+\frac{\eta ^{2}}{4}\phi &=&\left\{ 
\begin{array}{l}
\left( \nu +1/2\right) \phi ,\qquad \eta >0 \\ 
\\ 
\left[ \nu +1/2+s(\omega )\right] \phi ,\qquad \eta <0
\end{array}
\right.  \nonumber \\
&&  \label{8f} \\
-\frac{d^{2}\chi }{d\eta ^{2}}+\frac{\eta ^{2}}{4}\chi &=&\left\{ 
\begin{array}{l}
\left[ \nu +1/2+s(\omega )\right] \chi ,\qquad \eta >0 \\ 
\\ 
\left( \nu +1/2\right) \chi ,\qquad \eta <0
\end{array}
\right.  \nonumber
\end{eqnarray}
\noindent The second-order differential equations (\ref{8f}) have the form 
\begin{equation}
y^{\prime \prime }(z)-\left( \frac{z^{2}}{4}+a\right) y(z)=0,  \label{eq9}
\end{equation}
whose solution is a parabolic cylinder function \cite{abr}. The solutions $%
D_{-a-1/2}(z)$ and $D_{-a-1/2}(-z)$ are linearly independent unless $%
n=-a-1/2 $ is a nonnegative integer. In that special circumstance $D_{n}(z)$
has the peculiar property that $D_{n}(-z)=(-1)^{n}D_{n}(z)$, and it is
proportional to $\exp \left( -z^{2}/4\right) H_{n}(z/\sqrt{2})$, where $%
H_{n}(z)$ is a Hermite polynomial. The solutions of (\ref{8f}) do not
exhibit this parity property so that we should not expect nonnegative
integer values for $\nu $. The physically acceptable solutions for bound
states must vanish in the asymptotic region $|\eta |\rightarrow \infty $ and
are expressed as

\begin{eqnarray}
\phi &=&\theta (-\eta )C^{\left( -\right) }D_{\nu +s(\omega )}(-\eta
)+\theta (+\eta )C^{\left( +\right) }D_{\nu }(\eta )  \nonumber \\
&&  \label{eq10} \\
\chi &=&\theta (-\eta )D^{\left( -\right) }D_{\nu }(-\eta )+\theta (+\eta
)D^{\left( +\right) }D_{\nu +s(\omega )}(\eta )  \nonumber
\end{eqnarray}

\noindent where $C^{\left( \pm \right) }$ and $D^{\left( \pm \right) }$ are
normalization constants and $\theta (\eta )$ is the Heaviside function.
Substituting the solutions (\ref{eq10}) into the Dirac equation (\ref{eq8b})
and making use of the recurrence formulas

\begin{eqnarray}
\frac{d}{dz}D_{\nu }(z)-\frac{z}{2}D_{\nu }(z)+D_{\nu +1}(z) &=&0  \nonumber
\\
&&  \label{eq22} \\
\frac{d}{dz}D_{\nu }(z)+\frac{z}{2}D_{\nu }(z)-\nu D_{\nu -1}(z) &=&0 
\nonumber
\end{eqnarray}

\noindent one has as a result

\begin{eqnarray}
\left[ \frac{C^{\left( +\right) }}{D^{\left( +\right) }}\right] ^{2} &=&%
\frac{E+mc^{2}}{E-mc^{2}}\times \left\{ 
\begin{array}{c}
-\left( \nu +1\right) ,\qquad \omega >0 \\ 
-1/\nu ,\qquad \omega <0
\end{array}
\right.  \nonumber \\
&&  \label{22a} \\
\left[ \frac{C^{\left( -\right) }}{D^{\left( -\right) }}\right] ^{2} &=&%
\frac{E+mc^{2}}{E-mc^{2}}\times \left\{ 
\begin{array}{c}
-1/\left( \nu +1\right) ,\qquad \omega >0 \\ 
-\nu ,\qquad \omega <0
\end{array}
\right.  \nonumber
\end{eqnarray}

\noindent The continuity of the wavefunctions (\ref{eq10}) at $\eta =0$
furnishes

\begin{equation}
\frac{C^{\left( +\right) }}{D^{\left( +\right) }}=\frac{C^{\left( -\right) }%
}{D^{\left( -\right) }}\left[ \frac{D_{\nu +s(\omega )}\left( 0\right) }{%
D_{\nu }\left( 0\right) }\right] ^{2}  \label{22b}
\end{equation}

\noindent Together, (\ref{22a}) and (\ref{22b}) lead to the quantization
condition

\begin{equation}
D_{\nu +\left[ 1+s(\omega )\right] /2}\left( 0\right) \pm \sqrt{\pm \left[
\nu +(1+s(\omega ))/2\right] }D_{\nu -\left[ 1-s(\omega ))\right] /2}\left(
0\right) =0  \label{eq23}
\end{equation}

\noindent This last result combined with (\ref{eq8f}) \noindent shows that
the use of the minus sign inside the radical demands that $-\rho <\nu <-1$
for $\omega >0$, and $-\rho <\nu <0$ for $\omega <0$, here $\rho $ stands
for $1+m^{2}c^{4}/2\hslash |\omega |mc^{2}$. If, on the other side, one uses
the plus sign then $\nu >-1$ for $\omega >0$, and $\nu >0$ for $\omega <0$.

The numerical computation of (\ref{eq23}), which can be done easily with a
symbolic algebra program, reveals no solutions for $\nu <-1$. On the other
side, for $\nu >-1$ an infinite sequence of allowed values of $\nu $ are
found. The twenty lowest states for $\omega >0$ are given in Table 1. The
allowed values of $\nu $ for $\omega <0$ are those ones for $\omega >0$
added by the unit. By inspection of Table 1 one sees that for $\nu
\rightarrow \infty $ their values show a tendency to half-integer numbers.
The energy levels are obtained by inserting those allowed values of $\nu $
in (\ref{eq8f}):

\begin{equation}
E=\pm \sqrt{m^{2}c^{4}+2\hslash \left| \omega \right| mc^{2}\left\{ \nu
+\left[ 1+s(\omega )\right] /2\right\} }  \label{eq25}
\end{equation}

\noindent \noindent One should realize that the energy levels are
symmetrical about $E=0$. It means that the potential couples to the
positive-energy component of the spinor in the same way it couples to the
negative-energy component. In other words, this sort of potential couples to
the mass of the fermion instead of its charge so that there is no atmosphere
for the production of particle-antiparticle pairs. No matter the intensity
of the coupling parameter ($\omega $), the positive- and the negative-energy
solutions never meet. There is always an energy gap greater or equal to $%
2mc^{2}$, thus there is no room for transitions from positive- to
negative-energy solutions. This all means that Klein\'{}s paradox does not
come to the scenario.

It is noticeable in (\ref{eq10}) a somewhat left-right symmetry involving $%
\phi $ and $\chi $, such a symmetry is not exact inasmuch as $C^{\left( \pm
\right) }\neq $ $D^{\left( \mp \right) }$. The upper and lower components of
the spinor have the same number of zeros, or nodes, and as a consequence of
such a quasi-symmetry the zeros of $\phi $ and $\chi $ exhibit, of course,
an exact left-right symmetry. In whatever manner Eq. (\ref{22a})-(\ref{22b}%
), and (\ref{eq23}) of course, are invariant under the change $E\rightarrow
-E$ with the proviso that the normalization constants transforms as $%
|C^{\left( \pm \right) }|\longleftrightarrow |D^{\left( \mp \right) }|$,
hence one can conclude that $|\phi (\pm \eta )|\longleftrightarrow |\chi
(\mp \eta )|$ in such a way that $|\psi (\pm \eta )|\rightarrow |\psi (\mp
\eta )|$. If one recalls that the allowed values of $\nu $ for $\omega >0$
must be replaced by $\nu +1$ for $\omega <0$, a quasi-symmetry $|\phi (\pm
\eta )|\rightarrow |\phi (\mp \eta )|$ and $|\chi (\pm \eta )|\rightarrow
|\chi (\mp \eta )|$ can also be observed in (\ref{eq10}) under the
transformation $\omega \rightarrow -\omega $. This symmetry is not exact
because neither $|C^{(\pm )}|\rightarrow |C^{\left( \mp \right) }|$ nor $%
|D^{(\pm )}|\rightarrow |D^{\left( \mp \right) }|$. Figures 1--4 illustrate
the behaviour of $|\phi |^{2}$, $|\chi |^{2}$ and $|\psi |^{2}=|\phi
|^{2}+|\chi |^{2}$ for the ground-state contemplating all the possibilities
of signs of $E$ and $\omega $. Comparison of these Figures shows that $|\psi
|$ tends to concentrate at the left (right) region when $E$ and $\omega $
have equal (different) signs. It is also evident that $|\phi |$ is larger
(smaller) than $|\chi |$ when $E>0$ ($E<0$). Note that $|\phi |$ always
tends to concentrate at the left when $\omega >0$ ($\omega <0$). Figures 5
and Figure 6 illustrate the same things as Figure 1 for the first- and
second-excited-states. Note from all these Figures that the probability
position density has no nodes and the number of nodes of $\phi $, which is
equal to the one of $\chi $, increases with $\nu $. Furthermore, $|\phi |$
is larger (smaller) than $|\chi |$ when $E>0$ ($E<0$), independently of the
sign of $\omega $, as might be expected.

In addition to their intrinsic importance, the above conclusions renders a
contrast to the result found in \cite{mck}. There are bound-state solutions
for fermions interacting by a pseudoscalar Coulomb potential in 1+1
dimensions, to tell the truth there is confinement, notwithstanding the
spinor is not an eigenfunction of the parity operator. Therefore, the
quadri-dimensional version of this problem requires clarification for the
absence of bounded solutions. One might ponder that the underlying reason is
the way the spinors are affected by the behavior of the potentials at the
origin as well as at infinity because this is the radical difference between
the potentials in those two dissimilar worlds. In order to clarify this
point let us consider (\ref{eq8b}) for a general pseudoscalar potential.
Those equations involve the coupling between the upper and lower components
of the spinor. This coupling can be formally eliminated when these equations
are written as second-order differential equations:

\begin{equation}
-\frac{\hslash ^{2}}{2m}\frac{d^{2}\Phi }{dx^{2}}+V_{eff}^{\pm }\Phi
=E_{eff}\Phi  \label{30}
\end{equation}

\noindent where $\Phi $ refers to $\phi $ or $\chi $ and 
\begin{eqnarray}
E_{eff} &=&\frac{E^{2}-m^{2}c^{4}}{2mc^{2}}  \nonumber \\
&&  \label{30a} \\
V_{eff}^{\pm } &=&\frac{V_{p}^{2}}{2mc^{2}}\pm \frac{\hslash V_{p}^{\prime }%
}{2mc}, 
\begin{array}{c}
\qquad +\text{\textrm{for }}\phi \\ 
\qquad -\text{\textrm{\ for }}\chi
\end{array}
\nonumber
\end{eqnarray}

\bigskip \noindent \noindent \noindent \noindent these last results show
that the solution for this class of problem consists in searching for
bounded solutions for two Schr\"{o}dinger equations. It should not be
forgotten, though, that the equations for $\phi $ or $\chi $ are not indeed
independent because the effective eigenvalue, $E_{eff}$, appears in both
equations. Therefore, one has to search for bound-state solutions for $%
V_{eff}^{\pm }$ with a common eigenvalue. Now let us consider a
nonconserving-parity pseudoscalar potential in the form $V=\mu |x|^{\delta }$%
, then the effective potential becomes

\begin{equation}
V_{eff}^{\pm }=\frac{\mu ^{2}}{2mc^{2}}|x|^{2\delta }\pm s(x)\frac{\hslash
\mu \delta }{2mc}|x|^{\delta -1}  \label{32}
\end{equation}

\noindent Firstly let us consider the case $\mu >0$. When $\delta >0$ the
effective potential goes to infinity as $|x|\rightarrow \infty $ and it is
finite at the origin for $\delta \geq 1$ whereas for $0<\delta <1$ it is has
a singularity given by $\pm s(x)\mu \delta /2mc|x|^{1-\delta }$, implying
for $x>0$ in a potential-well structure for $V_{eff}^{+}$ and an attractive
potential less singular than $-1/8mx^{2}$ for $V_{eff}^{-}$ . The roles of $%
V_{eff}^{+}$ and $V_{eff}^{-}$ are changed for $x<0.$ Therefore, for $\delta
>0$ the power potential leads to effective potentials fulfilling the
conditions to furnish discrete spectra. On the other hand, when $\delta <0$
the effective potential vanishes as $|x|\rightarrow \infty $ and $%
V_{eff}^{-} $ ($V_{eff}^{+}$) is always repulsive at the origin for $x>0$ ($%
x<0$) in such a way that it is repulsive everywhere. Therefore, for $\delta
<0$ the power-law potential does not lead to bound-state solutions. Note
that these conclusions, either for $\delta >0$ or $\delta <0$, are
independent of the sign of $\mu $ on the condition that one has to change $%
V_{eff}^{\pm }$ to $V_{eff}^{\mp }$ if one changes the sign of $\mu $. If
one considers the conserving-parity potential $V=\mu x^{\delta }$ one obtains

\begin{equation}
V_{eff}^{\pm }=\frac{\mu ^{2}}{2mc^{2}}x^{2\delta }\pm \frac{\hslash \mu
\delta }{2mc}x^{\delta -1}  \label{33}
\end{equation}

\noindent and a similar analyzes also permits us to conclude that only when $%
\delta >0$ one could expect to find bounded solutions.

Now then, we return to focus our attention on the linear potential. Both
cases, if nonconserving- or conserving-parity, present potential-well
structures doing what is required to supply purely discrete spectra with
infinite sequences of eigenvalues, in other words they are confining
potentials, as it does every pseudoscalar power-law potential with $\delta
>0 $. The nonconserving-parity potential $V=m\omega c|x|$ gives rise to the
effective potential $V_{eff}^{\pm }=m\omega ^{2}x^{2}/2\pm s(x)\hslash
\omega /2$, an asymmetric potential with a discontinuity at $x=0$ equal to $%
\hslash \omega $. Now one can conceive why $|\psi |$ and $|\phi |$ are more
concentrated at the left region in the case $E>0$ and $\omega >0$. This
behaviour is precisely what one would expect for such an asymmetric
potential. On the other hand, the change $\omega $ $\rightarrow $ $-\omega $
results in $V_{eff}^{\pm }\rightarrow V_{eff}^{\mp }$ and one should expect
that $|\psi |$ and $|\phi |$ be more concentrated at the right side. In
addition, when the energy is sufficiently large ($\nu $ large) compared to $%
\hslash \omega $ the discontinuity of the effective potential at the origin
becomes unimportant and the fermion effectively undergo the influence of the
average effective potential ($m\omega ^{2}x^{2}/2$) maintaining their levels
(in the sense of $E_{eff}$) nearly equally spaced. This interpretation is
reinforced by observing Figures 1, 5 and 6, there one can see the propensity
of $|\psi |$ to be closer to the origin for larger energies. As a matter of
fact, a numerical calculation of the expectation value of $\eta $ (with $%
m=\omega =c=\hbar =1$) furnishes $-0.505,-0.121$ and $-0.084$ for the
ground-, first-excited and second-excited-states, respectively.

The conserving-parity potential $V=m\omega cx$ (the two-dimensional
generalized Dirac oscillator) is escorted by the effective potential $%
V_{eff}^{\pm }=m\omega ^{2}x^{2}/2\pm \hslash \omega /2$, which in addition
to be continuous at the origin is even under $x\rightarrow -x$. In this case
Eq. (\ref{8f}) for $\eta <0$ must be disregarded and the equations valid for 
$\eta >0$ must also be true for $\eta <0$, so the upper and lower components
of the Dirac spinor are given by 
\begin{eqnarray}
\phi &=&\theta (-\eta )C^{\left( -\right) }D_{\nu }(-\eta )+\theta (+\eta
)C^{\left( +\right) }D_{\nu }(\eta )  \nonumber \\
&&  \label{34} \\
\chi &=&\theta (-\eta )D^{\left( -\right) }D_{\nu +s(\omega )}(-\eta
)+\theta (+\eta )D^{\left( +\right) }D_{\nu +s(\omega )}(\eta )  \nonumber
\end{eqnarray}
and the eigenvalues are even expressed by (\ref{eq25}), nevertheless the
boundary conditions implies that $C^{\left( -\right) }=C^{\left( +\right) }$
and $D^{\left( -\right) }=D^{\left( +\right) }$, with 
\begin{equation}
\left[ \frac{C^{\left( +\right) }}{D^{\left( +\right) }}\right] ^{2}=\frac{%
E+mc^{2}}{E-mc^{2}}\times \left\{ 
\begin{array}{c}
-\left( \nu +1\right) ,\qquad \omega >0 \\ 
-1/\nu ,\qquad \omega <0
\end{array}
\right.  \label{36}
\end{equation}
In addition, the boundary conditions implies, for $\omega >0$, that $D_{\nu
}(0)=0$ with allowed values $\nu =1,3,5,\ldots $ and $D_{\nu }^{\prime
\,}(0)=0$ with allowed values $\nu =0,2,4,\ldots $ In this case $D_{\nu
}(-\eta )=(-1)^{\nu }D_{\nu }(\eta )$, and $D_{\nu }(\eta )$ is proportional
to $\exp \left( -\eta ^{2}/4\right) H_{\nu }(\eta /\sqrt{2})$, where $H_{\nu
}(\eta )$ is a Hermite polynomial (for $\omega <0$ one has to change the
allowed values of $\nu $ by $\nu +1$). The upper and lower components of the
spinor have definite and opposite parities under $\eta \rightarrow -\eta $
and, since the Hermite polynomial $H_{\nu }(\eta )$ has $\nu $ different
zeros, the number of nodes of $\chi $ is the number of nodes of $\phi $ plus
(minus) one for $\omega >0$ ($\omega <0$). This fact can be better
understood by keeping in mind that $V_{eff}^{\pm }=V_{eff}^{\mp }\pm \hslash
\omega $. The energy levels (in the sense of $E_{eff}$) are always equally
spaced because $\phi $ as much as $\chi $ are affected by continuous and
symmetric harmonic potentials. The symmetry already remarked for the
potential $V=m\omega c|x|$ translates into $|C^{\left( +\right)
}|\longleftrightarrow |D^{\left( +\right) }|$ and $|\phi (\eta
)|\longleftrightarrow |\chi (\eta )|$, under the change $E\rightarrow -E$,
in such a manner that $|\psi |$ is invariant. Also a quasi-symmetry is
present under $\omega \rightarrow -\omega $: $|\phi (\eta
)|\longleftrightarrow |\chi (\eta )|$. In this case the symmetry is not
perfect because neither $|C^{(+)}|\rightarrow |D^{(+)}|$ nor $%
|D^{(+)}|\rightarrow |C^{(+)}|.$ All in all, $|\phi |$ is larger (smaller)
than $|\chi |$ when $E>0$ ($E<0$), independently of the sign of $\omega $.

For short, in one-plus-one dimensions, we have succeed in searching
bound-state solutions for the Coulomb potential as well as for the
generalized Dirac oscillator. These problems are somehow similar as long as
they are affected by effective harmonic potentials and they, in fact,
possess the same formal expression for the eigenvalues. Nonetheless, the
quantum numbers and their related eigenfunctions differ because those
different problems are subject to distinct boundary conditions

As stated in the third paragraph of this work, the anomalous magnetic-like
interaction in the four-dimensional world turns into a pseudoscalar
interaction in the two-dimensional world. The anomalous magnetic interaction
has the form $-i\mu \beta \vec{\alpha}.\vec{\nabla}\phi (r)$, where $\mu $
is the anomalous magnetic moment in units of the Bohr magneton and $\phi $
is the electric potential, \textit{i.e.}, the time component of a vector
potential \cite{tha}. In one-plus-one dimensions the anomalous magnetic
interaction turns into $\sigma _{2}\mu \phi ^{\prime }$, then one might
suppose that the pseudoscalar Coulomb potential is due to an electric
potential proportional to $s(x)x^{2}/2$. The oddness of the electric
potential (under $x\rightarrow -x$) does not present problem to the
confinement of a fermion because its effective mass, an $x$-dependent mass
which always increases as the fermion goes away from the origin, depends on $%
\left( \phi ^{\prime \,}\right) ^{2}$, an result independent of the sign of $%
s(x)$ \cite{asc2}-\cite{asc3}.

A word should be said about the role of the $\pi $-meson field. The
Lagrangian density describing the pion-nucleon interaction $\mathcal{L}%
=-i\lambda \overline{\psi }\gamma _{5}\psi \pi $ is parity-invariant because
the $\pi $-meson is a pseudoscalar field,\textit{\ i.e.}, $\pi (\vec{r}%
,t)\rightarrow -\pi (-\vec{r},t)$ under the parity transformation.
Nonetheless, the interaction matrix term present in the Dirac equation as
written by McKeon and Van Leeuwen \cite{mck}, $i\beta \gamma _{5}V(r)$,
supposed to be due to the $\pi $-meson field is not parity-invariant due to
the reason that the potential function, $V(r)$, is parity-invariant but the
potential matrix is not. This argument exposes the inadequacy of the
interaction term in the Dirac equation as proposed in \cite{mck}. Therefore,
any conundrum about the role of the $\pi $-meson should be consistently
presented by taking into account a quintessential parity-invariant
potential. Furthermore, in order to correspond to the physical reality one
should be aware that the massive $\pi $-meson field gives rise to a Yukawa
potential instead of a Coulomb potential.

The methodology of effective potentials prompted in this paper might be
extended for the four-dimensional world. Such a methodology seems more
difficult there because the upper and lower components of the Dirac spinor
involve not only radial functions but also angular functions (the
two-component spinor spherical harmonics). Probably, we have to deal with
four coupled second-order differential equations. This task of carry the
Dirac equation through equivalent Sturm-Liouville problems is under way and
will be reported in due time.

\bigskip

\noindent \textbf{Acknowledgments}

The author wish to thank anonymous referees for very constructive remarks.
This work was supported in part by means of funds provided by CNPq and
FAPESP.

\smallskip \pagebreak

\bigskip \pagebreak

\noindent \textbf{Figure captions}

\smallskip

Figure 1 - $|\phi |^{2}$ (full thin line), $|\chi |^{2}$ (dashed line) and $%
|\phi |^{2}+|\chi |^{2}$ (full thick line), corresponding to \emph{positive}%
-ground-state energy for the potential $V=m\omega c|x|$ with $\omega >0$ and 
$m=|\omega |=c=\hslash =1$.

\smallskip

Figure 2 - The same as in Figure 1 with $\omega <0$.

\smallskip

Figure 3 - The same as in Figure 1 corresponding to \emph{negative}%
-ground-state energy ($\omega >0$).

\medskip

Figure 4 - The same as in Figure 1 corresponding to \emph{negative}%
-ground-state energy with $\omega <0$.

\smallskip

Figure 5 - The same as in Figure 1 corresponding to \emph{positive}%
-first--excited-state energy ($\omega >0$).

\smallskip

Figure 6 - The same as in Figure 1 corresponding to \emph{positive}%
-second-excited-state energy ($\omega >0$).

\bigskip \pagebreak

\begin{table}[tbp]
\caption{The lowest allowed values of $\nu $, for $\omega >0$, such that Eq.
(21) is satisfied.}
\label{t1}
\begin{center}
\begin{tabular}{|c|c|}
\hline\hline
$\nu$ \textrm{for plus sign in front of the radical} & $\nu$ \textrm{for
minus sign in front of the radical} \\ \hline
-0.654541 & 0.5485571 \\ 
1.468573 & 2.522295 \\ 
3.482395 & 4.514350 \\ 
5.487785 & 6.510563 \\ 
7.490650 & 8.508353 \\ 
9.492428 & 10.506906 \\ 
11.493638 & 12.505886 \\ 
13.494514 & 14.505129 \\ 
15.495179 & 16.504543 \\ 
17.495700 & 18.504078 \\ \hline\hline
\end{tabular}
\end{center}
\end{table}


\begin{thebibliography}{99}
\bibitem{col1}  S. Coleman, R. Jackiw and L. Susskind, Ann. Phys. (N.Y.) 93
(1975) 267.

\bibitem{col2}  S. Coleman, Ann. Phys. (N.Y.) 101 (1976) 239.

\bibitem{fro}  J. Fr\"{o}hlich and E. Seiler, Helv. Phys. Acta 49 (1976) 889.

\bibitem{cap}  A.Z. Capri and R. Ferrari, Can. J. Phys. 63 (1985) 1029.

\bibitem{gal}  H. Gali\'{c}, Am. J. Phys. 56 (1988) 312.

\bibitem{hoo}  G.\'{}t Hooft, Nucl. Phys. B\textbf{\ }75 (1974) 461.

\bibitem{kog}  J. Kogut and L. Susskind, Phys. Rev. D 9 (1974) 3501.

\bibitem{bha}  R.S. Bhalerao and B. Ram, Am. J. Phys. 69 (2001) 817.

\bibitem{cas}  A. S. de Castro, Am. J. Phys. 70 (2002) 450.

\bibitem{cav}  R.M. Cavalcanti, Am. J. Phys. 70 (2002) 451.

\bibitem{hil}  J.R. Hiller, Am. J. Phys. 70 (2002) 522.

\bibitem{luc}  W. Lucha, F.F. Sch\"{o}berl and D. Gromes, Phys. Rep. 200
(1991) 127 and references therein.

\bibitem{asc1}  A.S. de Castro, Phys. Lett. A 305 (2002) 100.

\bibitem{mck}  D.G.C. McKeon and G. Van Leeuwen, Mod. Phys. Lett. A 17
(2002) 1961.

\bibitem{str}  P. Strange, Relativistic Quantum Mechanics, Cambridge
University Press, Cambridge, 1998.

\bibitem{asc2}  A.S. de Castro and W.G. Pereira, Phys. Lett. A 308 (2003)
131.

\bibitem{asc3}  A.S. de Castro, Phys. Lett. A 309 (2003) 340.

\bibitem{ms}  M. Moshinsky and A. Szczepaniak, J.~Phys.~A 22 (1989) L817.

\bibitem{nt}  Y. Nogami and F.M. Toyama, Can. J. Phys. 74 (1996) 114.

\bibitem{tplus}  F.M. Toyama, Y. Nogami and F.A.B. Coutinho, J. Phys. A 30
(1997) 2585.

\bibitem{tn}  F.M. Toyama and Y. Nogami, Phys. Rev. A 59 (1999) 1056.

\bibitem{sg}  R. Szmytkowski and M. Gruchowski, J. Phys. A 34 (2001) 4991.

\bibitem{pa}  M.H. Pacheco, R. Landim and C.A.S. Almeida, Phys. Lett. A 311
(2003) 93.

\bibitem{tha}  B. Thaller, The Dirac Equation (Springer-Verlag, Berlin,
1992).

\bibitem{abr}  M. Abramowitz and I. A. Stegun, Handbook of Mathematical
Functions (Dover, Toronto, 1965).
\end{thebibliography}
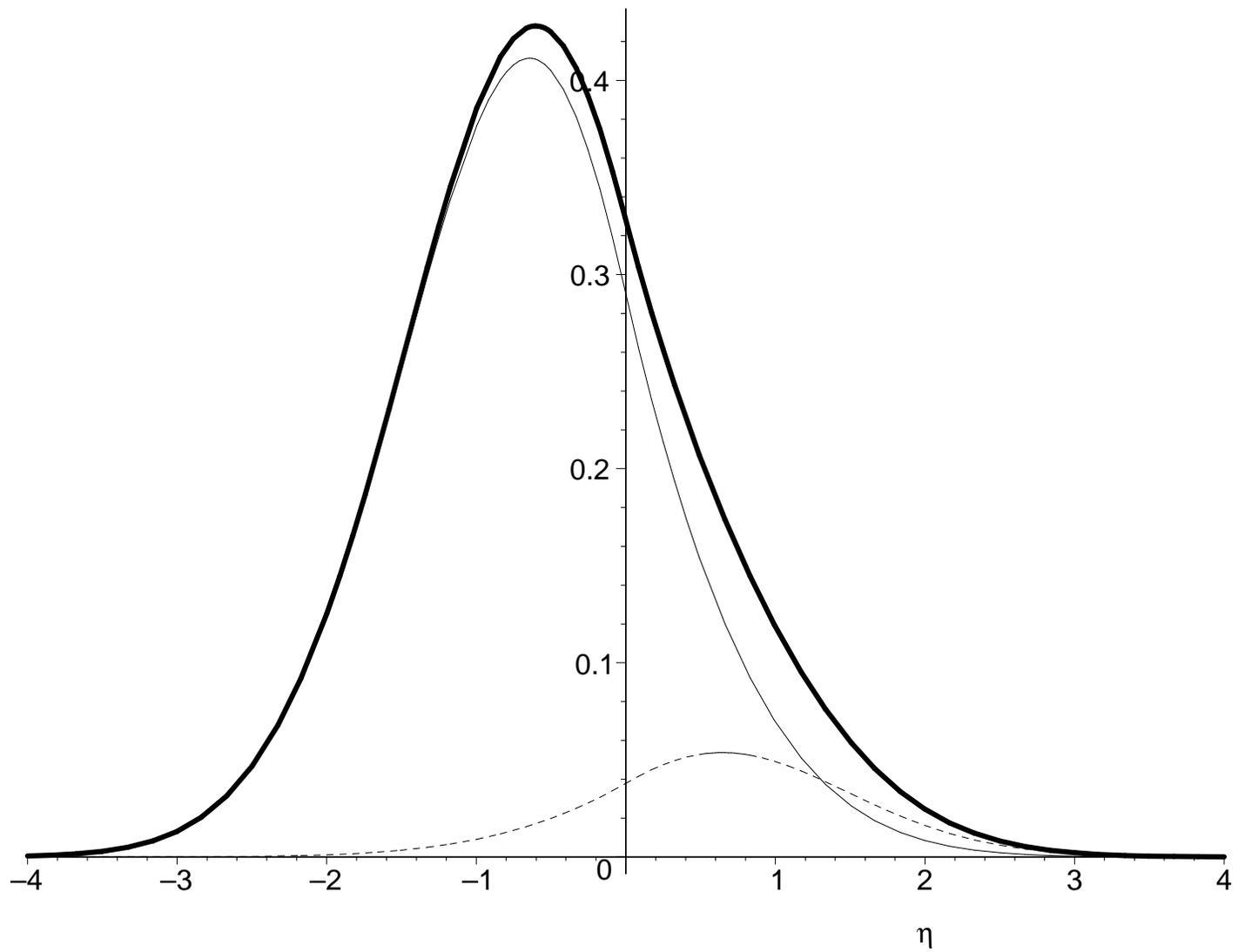
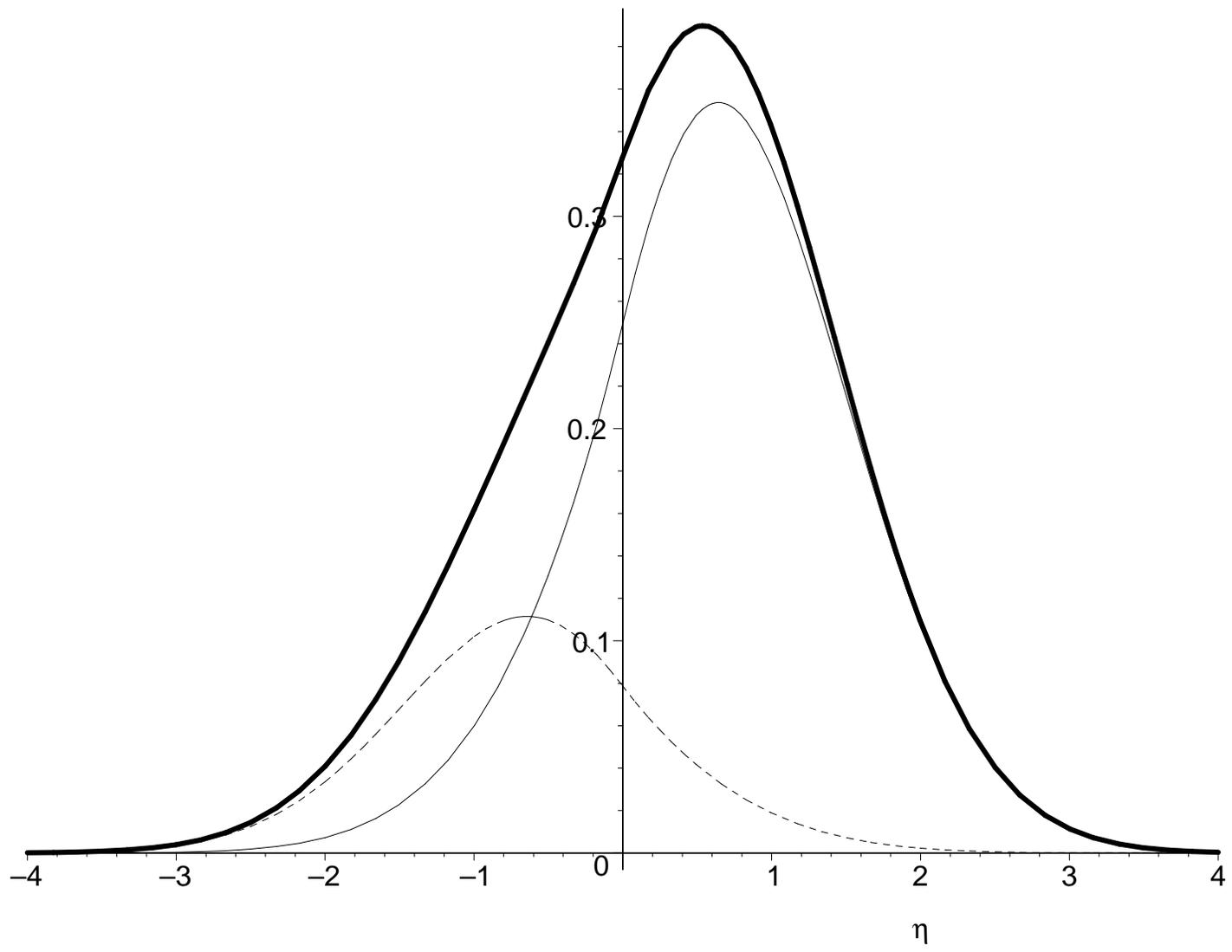
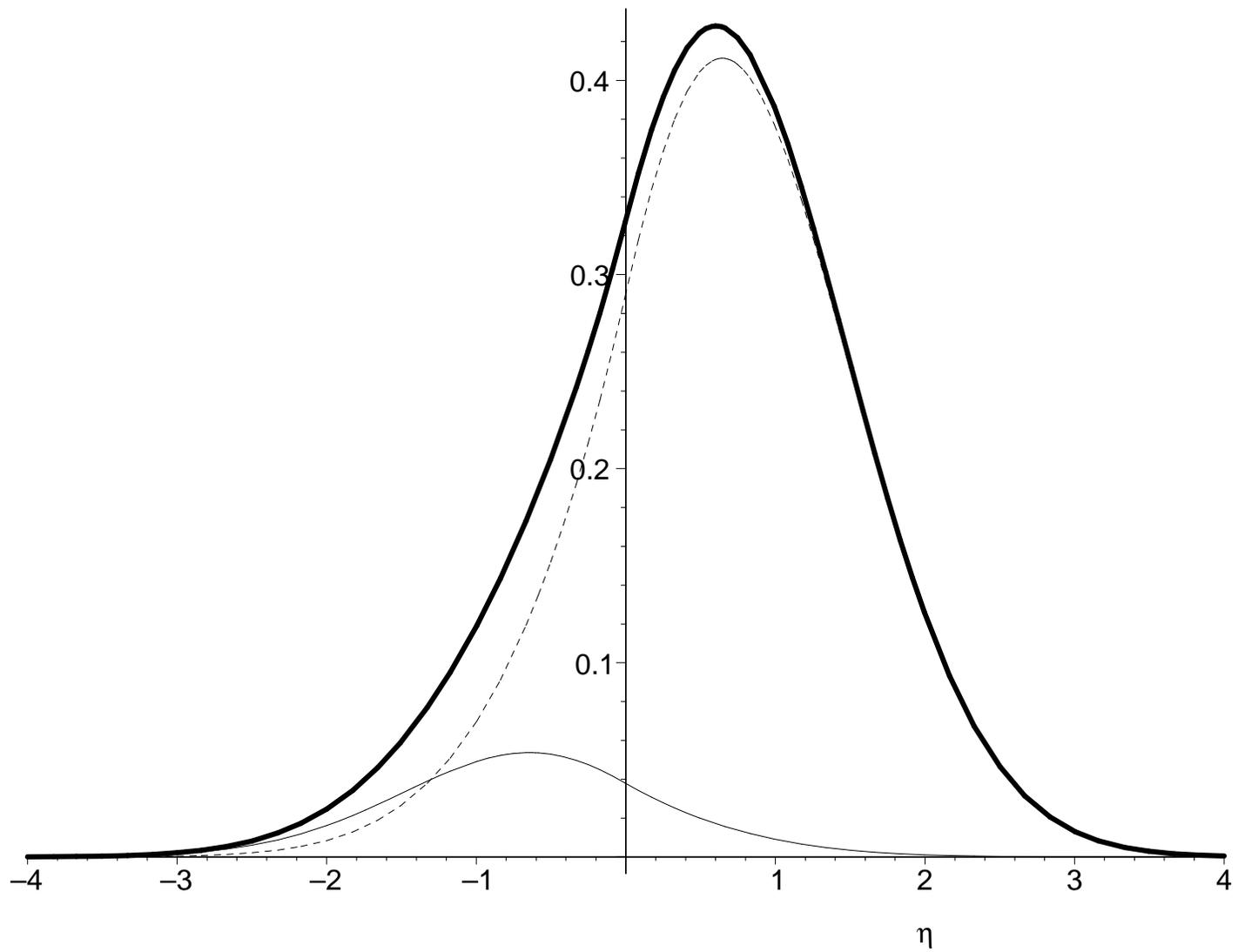
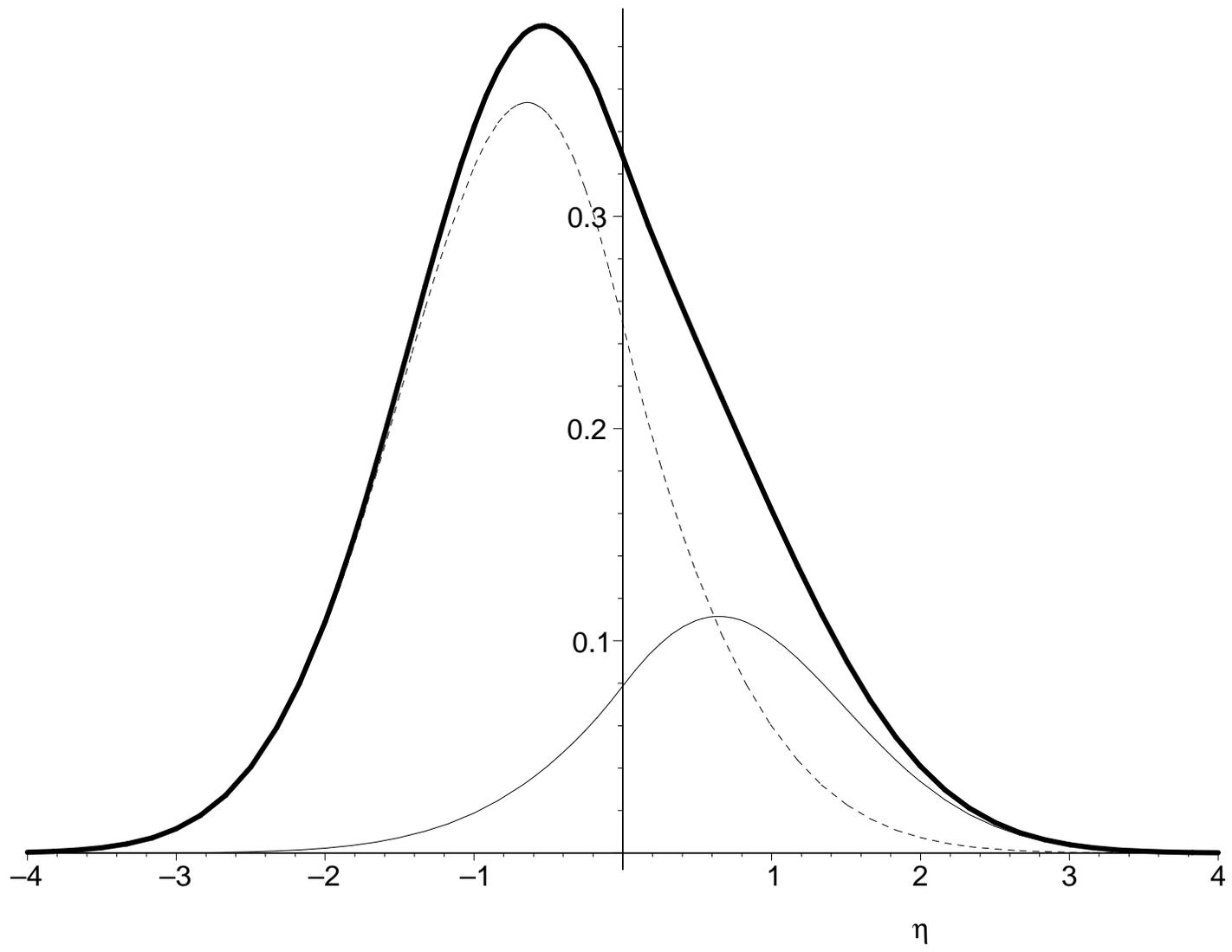
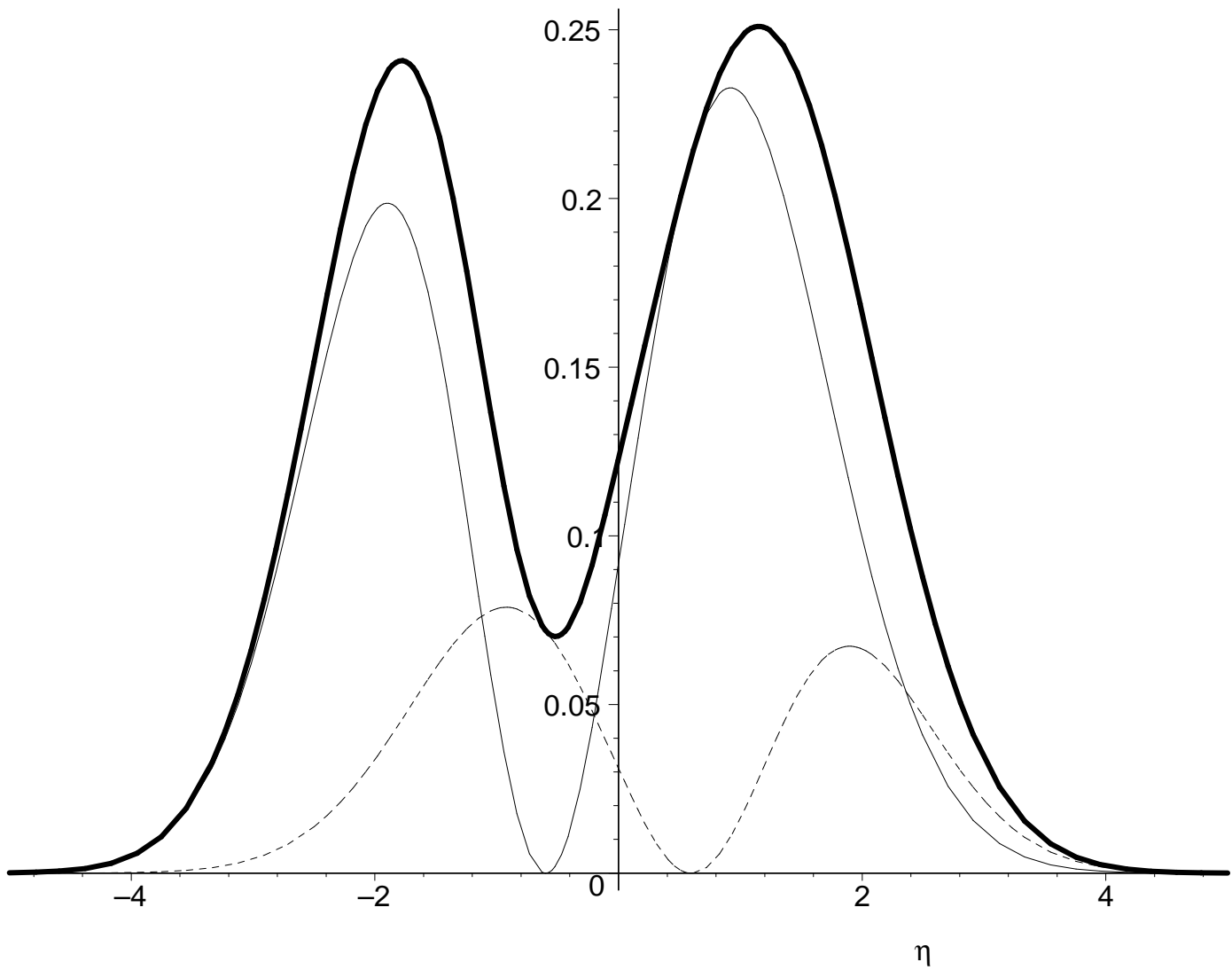
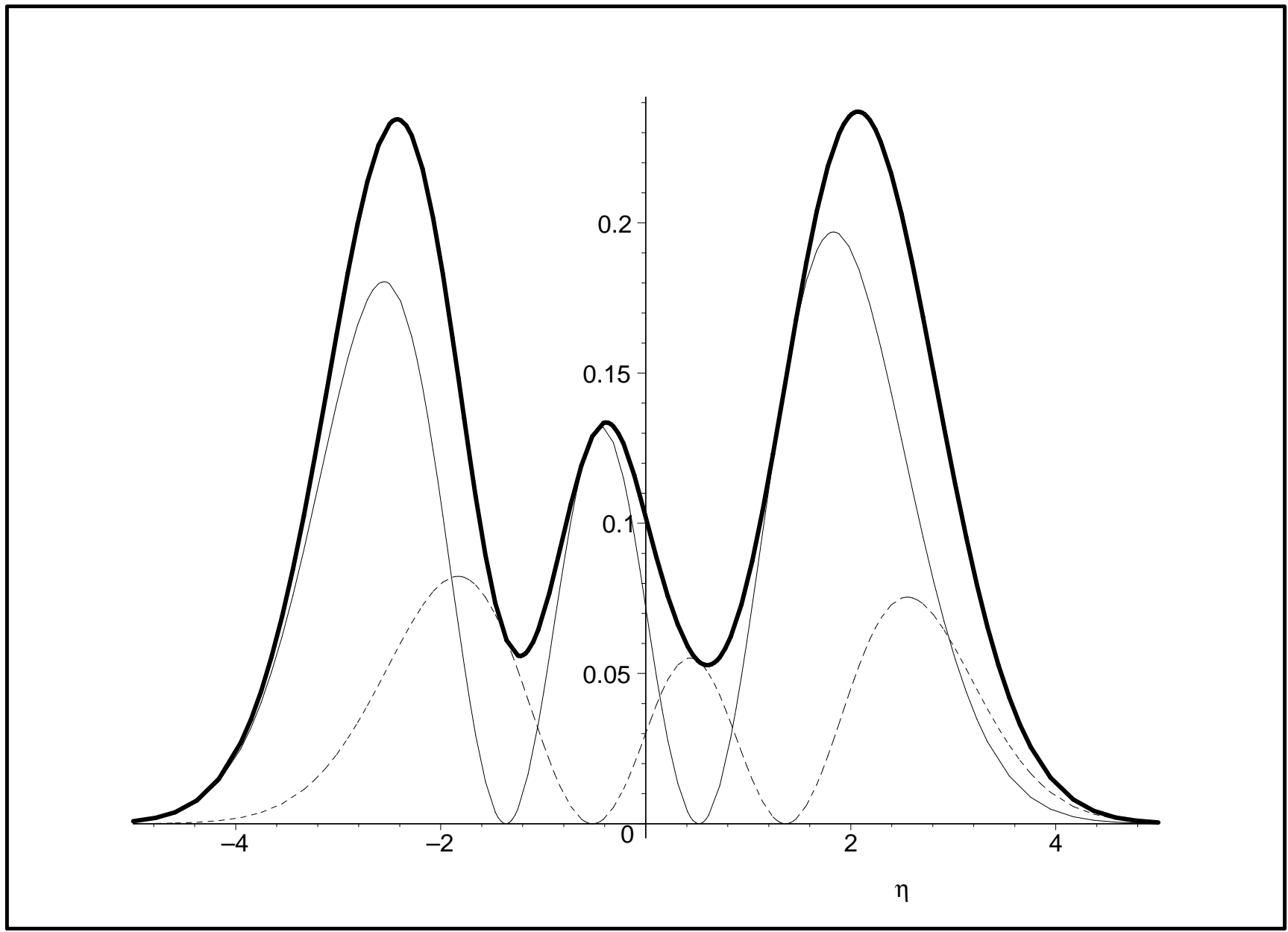
\end{document}